\documentclass[structabstract]{aa}

\usepackage{graphicx}
\usepackage{txfonts}
\usepackage{epsfig}
\usepackage{tabularx}

\begin{document}
\title{Effects of tidally enhanced stellar wind on the horizontal branch morphology of globular clusters}

\author{Zhen-xin Lei
  \inst{1,2,3}
    \and
  Xue-Fei Chen \inst{1,2}
  \and
  Feng-Hui Zhang \inst{1,2}
  \and
  Z. Han \inst{1,2}
}


\institute{National Astronomical Observatories/Yunnan Observatory,
  the Chinese Academy of Sciences, Kunming 650011, China\\
  \email{lzx2008@ynao.ac.cn; cxf@ynao.ac.cn;
  zhangfh@ynao.ac.cn; zhanwenhan@ynao.ac.cn}
  \and
  Key Laboratory for the Structure and Evolution of Celestial Objects, the Chinese Academy of Sciences, Kunming 650011, China
  \and
  University of the Chinese Academy of Sciences, Beijing 100049, China}

\date{Received ; accepted}


\abstract
{Metallicity is the first parameter to influence the horizontal branch (HB)
morphology of globular clusters (GCs). It has been found, however, that
some other parameters may also play an important role in affecting
the morphology. While the nature of these important parameters
remains unclear, they are believed to be likely correlated with wind mass-loss
of red giants, since this mass loss determines their subsequent
locations on the HB. Unfortunately,
the mass loss during the red giant stages of the stellar evolution
is poorly understood at present. }
{ The stellar winds of red giants may be tidally enhanced
by companion stars if they are in binary systems.
We investigate evolutionary consequences of red giants in binaries
by including tidally enhanced stellar winds, and examine the effects on the
HB morphology of GCs.}
{We used Eggleton's stellar evolution
code to study the binary evolution.
The tidally enhanced stellar-wind model of Tout \& Eggleton is
incorporated into this code,
where the tidal enhancement parameter, $B\rm_{w}$,
has various values (e.g., 10000 and 500) to examine
the dependency of the final results on this parameter.
A Monte Carlo simulation was performed to generate a
group of binary systems. The position of each primary star
on the HB in the Hertzsprung-Russell (H-R) diagram in this sample
is obtained through interpolations among the
constructed HB evolutionary tracks.  Finally, a synthetic HB in
the color-magnitude diagram is obtained by transforming
the effective temperature and luminosity of each primary star on
the HB into $B-V$ colors and absolute magnitude. }
{We find that red, blue, and extreme horizontal branch stars are all
produced under the effects of tidally enhanced stellar wind without any
additional assumptions on the mass-loss dispersion.  Furthermore, the
horizontal branch morphology is found to be insensitive to the tidal
enhancement parameter, $B\rm_{w}$.   We compare our theoretical results
with the observed horizontal branch morphology of globular cluster NGC
2808, and find that the basic morphology of the horizontal branch
can be well reproduced.
The number of blue horizontal branch stars in our calculations,
however, is lower than that of NGC 2808.}
{}

\keywords{Stars: horizontal-branch -- stars: mass-loss --
  Stars: binaries: general -- globular clusters: general}

\titlerunning{Effects of tidally-enhanced stellar wind on HB morphology.}
\authorrunning{Zhen-xin Lei et al. }

\maketitle

%

\section{Introduction}
For the first time, Hoyle \& Schwarzschild (1955) have identified
horizontal branch (HB) stars on the Hertzsprung-Russell (H-R)
diagram as the progeny of red giant branch (RGB) stars.  HB stars
in globular clusters (GCs) are low-mass stars, which burn helium
in their cores. They have different hydrogen-envelope masses, but
nearly the same core mass (Iben \& Rood, 1970).  HB stars play a
very important role in many aspects of astrophysics.  For instance,
we can use HB stars to obtain many GC parameters, such as
cluster age, helium abundance, cluster distance, etc.  HB stars are
also used to test stellar structure and evolution models (e.g., Renzini
\& Fusi Pecci, 1988; VandenBerg et al. 1996).  Hot HB stars and their
progeny are considered to be the main contributors of ultraviolet
(UV) excess emission found in early-type galaxies (e.g., Kilkenny et
al. 1997; Han, Podsiadlowski \& Lynas-Gray, 2007).

The color distribution of HB stars in GCs on the color-magnitude
diagram (CMD), which is called HB morphology, is very different
among GCs in the Milky Way.  The metallicity is
considered to be the most important parameter influencing the
HB morphology and is called the first parameter (see Sandage
\& Wallerstein, 1960).  Metal-rich GCs have a redder HB morphology,
in contrast, metal-poor GCs have a bluer morphology.  It has been found, however,
that metallicity cannot explain the whole HB morphology of GCs in the
Milky Way.  Some GCs exhibit similar metallicities, but
have very different HB morphologies (e.g., M3 and M13; Rey et al.
2001).  Furthermore, some metal-rich GCs show extreme HB (EHB)
stars (e.g., NGC 6388 and NGC 6441; Rich et al. 1997), which
cannot be explained by the effects of metallicity only.
This implies that there are some other key
factors at work in addition to the metalicity that influence the HB
morphology of GCs (for a recent review see Catelan,
2009).  These parameters are collectively known as the
``second parameter'' (2P).  Over the past decade, many 2P
candidates have been proposed, such as cluster age (Lee et al.
1994); internal rotation and helium mixing (Swigart, 1997);
helium self-enrichment (D'Antona et al. 2002; D'Antona et al.
2005; D'Antona \& Caloi, 2004, 2008); core density or
concentration (Fusi pecci et al. 1993; Buonanno et al.
1997); presence of a planetary system (Soker, 1998; Soker et al.
2000, 2001, 2007); cluster mass (Recio-Blanco et al. 2006), etc.
However, most of these 2P candidates are based on the
single-star evolution, and none of them can alone account for
the whole HB morphology in GCs.

The 2P problem in GCs is considered to be correlated with the
mass loss of the RGB stars,  since the position of
stars on the HB in GCs is determined by their envelope mass,
while the envelope mass depends on the mass loss of the
RGB stars.  Unfortunately, the physical mechanisms
of mass loss on the RGB are poorly understood at present (Willson,
2000; Dupree et al. 2009).  Therefore, to reproduce
different parts of the HB in GCs, most of the 2P candidates based on
single-star evolution
need to assume a
mass-loss dispersion on the RGB (e.g., Gaussian distribution of
mass for HB stars: Lee et al. 1994; D'Antona \& Caloi, 2004).
The assumption of mass loss dispersion, however, is completely
arbitrary and without physical justification.

Fusi pecci et al. (1993) and Buonanno et al. (1997) proposed
that dense and concentrated environment in GCs yields a
bluer and longer HB morphology.  These studies implied
that a  stellar interactions (e.g., binary interactions)
may affect the HB morphology by
enhancing the mass loss of the HB progenitors.
In a binary system, the primary star may
fill its Roche lobe on the RGB if the orbital period is short enough.
In doing so, the primary begins to steadily transfer its envelope mass to
the secondary star through Roche lobe overflow (RLOF),
or forms a common envelope (CE) if the
mass transfer is unstable.  Once the process of RLOF or CE
is complete,
the envelope mass of the primary star becomes very thin.  If
helium in the core is ignited, the primary star
settles on the EHB in the H-R diagram (Han et al. 2002, 2003).
However, this model can only produce EHB stars.
On the other hand, although some binary systems with longer orbital periods may
not fill their Roche lobe,
Tout \& Eggleton (1988) suggested that
the companion star could tidally enhance the stellar wind
of the red giant primary, and
this tidally enhanced stellar wind
during the binary evolution could explain the mass inversion (i.e.,
a more evolved star exhibits lower mass) observed in some RS
CVn binaries.  For this kind of situation, the primary can also lose
much of its envelope mass through stellar wind and be located on
a blue position of the HB after helium is ignited in its core.
Under the tidally enhanced stellar wind,
the envelope mass of HB stars is determined by the
separation of the two companion stars in the binary system.
With different initial binary orbital periods, the primary stars will
lose a different envelope mass on the RGB and then be located at
different places on the HB.

In this paper, we investigate the consequences of
tidally enhanced stellar winds during binary evolution
on enhancing the mass loss of red giants, with
accompanying effects on the horizontal branch morphology
of globular clusters.  The RLOF and CE processes
during binary evolution are not considered in this work.

The structure of this paper is as follows: in Section~2, we introduce
the method and the numerical code.  Our
results and comparison with observations are presented in Section~3.
The results are discussed in Section~4, and final
conclusions are drawn in Section~5.

\section{Methodology}

To investigate the effects of tidally enhanced
stellar winds during the binary evolution on
the HB morphology of GCs,
we incorporated the tidally enhanced stellar wind (described in Section 2.2 below)
into the Eggleton's stellar evolution code
to calculate the stellar mass and the helium core mass
of the primary at helium flash
(hereafter $M\rm_{HF}$ and $M\rm_{c,HF}$, respectively)
after experiencing mass loss on the RGB.
Then,  $M\rm_{HF}$, $M\rm_{c,HF}$, together with the time spent on the HB
were used to obtain the positions of the primary stars on the HB
\footnote{Since $M\rm_{HF}$ and $M\rm_{c,HF}$
are little changed from the RGB tip to zero-age HB, these two
parameters are used as the
same parameters for HB stars at zero-age HB to determine the
positions of HB stars in the H-R diagram.} in H-R diagram
by interpolating among the constructed HB evolutionary tracks,
which are described in Section~3.2.
Finally, we transformed the effective temperature and luminosity into
$B-V$ colors and absolute magnitude, $M\rm_{v}$,  and obtained the
synthetic HB morphology in CMD, which could then be directly compared
with observations.

\subsection{Stellar evolution code } \label{bozomath}

The original Eggleton stellar evolution code was developed four
decades ago (Eggleton, 1971, 1972, 1973).  New physics has been
added to the code by Han et al. (1994) and Pols et al.
(1995, 1998).  At present, the advanced stellar evolution code is being
applied extensively in the field of stellar evolution.

We here incorporate the tidally enhanced stellar wind (see Section
2.3 below) into the Eggleton stellar
evolution code to evolve the binary systems. In this code, we chose a
ratio of mixing length, $l$, to local pressure scale-height, $H\rm_{P}$, of
$\alpha=l/H\rm_{P}=2.0$.  The convective overshooting parameter in the
code, $\delta\rm_{ov}$, is assumed to be $0.12$ (Pols et al. 1997).  The
opacity used in our calculations is the same as that compiled by
Chen \& Tout (2007).  We obtain the initial hydrogen mass fraction,
$X$, using the relation $X$ = 0.76-3$Z$ (Pols et al. 1998), where
$Z$ is the metallicity.

\subsection{Tidally enhanced stellar wind } \label{bozomath}

Most binaries of RS CVn type consist of a red subgiant and an MS star.
It has been found, however, that some of these systems exhibit a
mass inversion phenomenon (e.g., the more evolved star has a lower mass)
before the RLOF takes place (e.g., Popper, \& Ulrich 1977; Popper, 1980).
Tout \& Eggleton (1988)
found that the standard mass-loss rate of red giants (e.g., Reimers
mass-loss rate; Reimers, 1975) is too low to yield the observed
mass inversion.  Tout \& Eggleton suggested that the secondary star may
tidally enhance the stellar wind of the primary star.
Since the torque due to tidal friction depends on $(R/R_{L})^{6}$
(Zahn, 1975; Campbell \& Papaloizou, 1983 ), they used the
following equation to describe the tidally enhanced stellar wind of
the red giant primary:
\begin{equation}
\dot{M}=-\eta4\times10^{-13}(RL/M)\{1+B\rm_{w}\times \rm min[{\it(R/R_{L})}^{6},
\rm 1/2^{6}]\},
\end{equation}
where $\eta$ is the Reimers mass-loss efficiency (Reimers, 1975), $R_{L}$ is the
radius of Roche lobe, and $B\rm_{w}$ is the efficiency of the tidal
enhancement of the stellar wind.  Here $R$, $L$, and $M$ are
in solar units.  Tout \& Eggleton introduced a saturation
for $R \ge 1/2R_{L}$ in the above expression, because it is expected
that the binary system is in complete corotation for $R/R_{L} \ge
0.5$. We added Equation (1) to the
Eggleton stellar evolution code to
study the effects of tidally enhanced stellar
wind during binary evolution on the HB morphology
of GCs.  The
results are given in Table~1 ( see Section~3.1).

Because the time that stars spent
on the HB is very short (e.g., $10^{8}$ years) relative to the
lifetime spent on the MS (e.g., $10^{10}$ years), the initial mass
of stars in a GC that resides on the HB is nearly the same (apparently,
this mass depends on the age of GC).
In our model calculations, the initial stellar mass
of the primary star was chosen to be 0.85$M_{\odot}$, and the
metallicity was set to $0.001$, which is a typical value for GCs.
This mass corresponds to a star age of about $11.6$ Gyr at the
RGB tip.

In our model calculations, we set
the mass ratio of primary-to-secondary to be $1.6$
(see the discussion in Section~4).
We used the equation of Bondi \& Hoyle (1944) to calculate the
accretion rate from the stellar wind of the primary star onto
the companion star.  We chose a typical stellar wind speed of
$15$~kms$^{-1}$ and assumed that the entire mass accreted by
the secondary star from stellar wind is retained within the binary
system.  The angular momentum that leaves the binary system
due to the stellar wind is attributed to the primary star.

\subsection{Initial binary samples} \label{bozomath}

To obtain the synthetic HB in CMD which can be directly
compared with observations,
we generated a group of binary systems.
The initial stellar masses of all primary stars were chosen to be
0.85$M_{\odot}$ and  the mass ratio of primary to secondary
was set to be 1.6 (see Section~2.2).
The initial binary orbital periods (or star separations) were
produced by Monte Carlo simulations.
The distribution of star separation in
the binary system was assumed to be uniform in $\log a$ space
(with $a$ being the star separation) and falls off smoothly for
low values of $a$ (see Han et al. 2003), that is,
\begin{equation}
a\cdot n(a)=\left\{
 \begin{array}{lc}
 \alpha_{\rm sep}(a/a_{\rm 0})^{\rm m}, & a\leq a_{\rm 0},\\
\alpha_{\rm sep}, & a_{\rm 0}<a<a_{\rm 1},\\
\end{array}\right.
\end{equation}
where $\alpha_{\rm sep}\approx0.07$, $a_{\rm 0}=10\,R_{\odot}$,
$a_{\rm 1}=5.75\times 10^{\rm 6}\,R_{\odot}=0.13\,{\rm pc}$, and
$m\approx1.2$.  This distribution implies that the number of wide
binaries per logarithmic interval is equal, and that about 50\% of
the stellar systems have orbital periods shorter than $100$~yr.

\section{Results}

\subsection{Stellar mass and helium core mass of the primary at helium flash}
\label{bozomath}

\begin{table}
  \small
  \begin{center}

    \begin{minipage}[]{90mm}
      \caption[]{Stellar mass and helium core mass at the stage of
        the helium flash (for $M\rm_{ZAMS}$=0.85$M_{\odot}$) for various
        initial orbital periods.  Here $Z$=0.001, $B\rm_{w}$=10000,
        $\eta$=0.25, and $q$=1.6. The age of the primary stars
        at helium flash is about 11.6 Gyr. }
    \end{minipage}\\

    \begin{tabularx}{80mm}{XXXX}
      \hline\noalign{\smallskip}
      log$P/\rm day$  \qquad& $M\rm_{HF}$($M_{\odot}$)     \qquad& $M\rm_{c,HF}$($M_{\odot}$)    \\
      \hline\noalign{\smallskip}
        3.1438$\rm^{a}$    & 0.4716      & 0.4706         \\
        3.2000             & 0.4807      & 0.4798         \\
        3.2300             & 0.4857      & 0.4849         \\
        3.2400             & 0.4874      & 0.4865         \\
        3.2500             & 0.4890      & 0.4875         \\
        3.2600$\rm^{b}$    & 0.4908      & 0.4886         \\
        3.2700             & 0.4942      & 0.4887         \\
        3.2800             & 0.5038      & 0.4887         \\
        3.3000             & 0.5323      & 0.4887         \\
        3.3200             & 0.5640      & 0.4887         \\
        3.3400             & 0.5943      & 0.4887         \\
        3.3600             & 0.6211      & 0.4886         \\
        3.3800             & 0.6439      & 0.4885         \\
        3.4000             & 0.6630      & 0.4884         \\
        3.4500             & 0.6977      & 0.4883         \\
        3.5000             & 0.7191      & 0.4882         \\
        4.0000             & 0.7540      & 0.4880         \\
        10.000             & 0.7543      & 0.4880         \\
        \hline\noalign{\smallskip}
      \end{tabularx}
    \end{center}

    $\rm^{a}$: The minimum initial orbital period of the binary required by the helium flash.

    $\rm^{b}$: The minimum initial orbital period of the binary above which the helium flash
    takes place at the RGB tip.
\end{table}

Table 1 gives $M\rm_{HF}$ and  $M\rm_{c,HF}$ of the primary stars
at the helium flash (after mass loss experienced while on the RGB) for
different initial binary orbital periods.
In these calculations, the Reimers mass-loss efficiency, $\eta$, was set
to be $0.25$, and the tidal enhancement efficiency, $B\rm_{w}$, was chosen to be
$10^4$ (Tout \& Eggleton, 1988). In the table, the columns from
left to right provide the initial orbital period of the binary system,
$M\rm_{HF}$, $M\rm_{c,HF}$,  respectively.

The first row of Table~1 gives the
the minimum orbital period (log$P/\rm day$) of the binary system
above which the helium flash may take place in
our calculations. For shorter periods \footnote{
If the binary orbital period is short enough to
make the primary star fill its Roche lobe
on the RGB,  a RLOF or CE process
will happen in this binary system. However, this is beyond
the scope of the present work.}, the primary star may lose too
much envelope mass through the tidally enhanced stellar wind, and
hence its helium core is too small to ignite helium. These
stars would evolve straight into helium white dwarfs (WDs), and they will
not experience a helium flash; in other words, they will follow a
WD cooling curve after the RGB phase.

With increasing initial orbital period, $P/\rm day$,
the mass loss experienced by the primary
star during the RGB phase decreases.
If $P/\rm day$ becomes long enough,
the tidally enhanced stellar wind becomes
unimportant and has little effect on
the amount of mass loss of the primary stars on the RGB
(one can see
that the primary stars with an orbital period of
log$P/\rm day\rm=4.0$ and log$P/\rm day\rm=10.0$ in Table~1
have nearly the same stellar mass at helium flash, which
means that they experience nearly the same amount of
mass loss on the RGB), because the
separation of the two companion stars becomes too large.
In this case, the primary star loses envelope
mass only by means of Reimers mass loss.

The helium flash takes place at the RGB tip for binary systems with
log$P/\rm day\rm\geq3.26$ (see Table~1).  For shorter orbital
periods, the primary star loses too much envelope mass
and  experiences a helium flash
at higher temperatures on the H-R diagram. This kind of flash is called {\em
hot flash} (Castellani \& Castellani, 1993; D'Cruz et al. 1996).
It has been suggested that there are two types of hot flashes.
The first type is {\em early hot flash}, for which the helium
flash occurs when the star crosses the H-R diagram toward
the WD cooling curves.  The chemical
composition of the envelope of early hot flashers is not changed
by the helium flash.  Stars that undergo an early hot flash are
located on the blue end of the canonical EHB of the H-R diagram
(e.g., see Fig.~9 in Brown et al. 2001).  The second type
is called {\em late hot flash}, for which the helium flash
occurs on a WD cooling curve (after the primary star experiences
a huge mass loss on the RGB).  In this type, the
helium flash {\em can} change the chemical composition of its
envelope by enhancing the helium and carbon abundance through
helium-flash mixing (Iben, 1976; Sweigart, 1997; Brown et al.
2001).  Therefore, the late hot flashers have significantly higher
effective temperatures on the HB, and they are fainter than the
canonical EHB stars in the CMD.

The late-hot-flash model has been used to
explain the existence of blue-hook stars found in several massive
GCs (Whitney et al. 1998; D'Cruz et al. 2000; Brown et al. 2001).
In our calculations, however, we did not consider the helium-flash
mixing process.  That is because the Eggleton stellar evolution
code cannot pass the helium-flash phase for low-mass stars.
Hence, we do not know how much helium and carbon will be
mixed in the envelope during the late-hot-flash phase.  For this
reason, all hot flashers in our calculation are treated as early
hot flashers, for which the envelope's chemical composition remains
unchanged.  This assumption will not change the number of
EHB stars produced in our models, and hence will not
influence our final results.

\subsection{HB evolutionary tracks}

To obtain the position of the primary stars on the HB in the H-R diagram, we
need to construct the HB evolutionary tracks.  However,
the Eggleton's stellar evolution code cannot pass the helium flash
stage for low-mass stars (e.g., $M/M_{\odot}\leq 2.0$;
this value is the function of metallicity).
For this reason, we are only able to construct the
zero-age HB (ZAHB) models using
stars with greater initial masses (e.g., 2.5$M_{\odot}$).
The drawback of this method is that the chemical profile of the
envelope in ZAHB models constructed by massive stars is different
from the one that an actual low-mass star would have had on the
ZAHB\footnote{However, with some effort,
we can also construct the ZAHB
models with chemical profiles similar to that of low-mass stars
at the tip of the RGB, but it is not very convenient.}.

For the above reasons, the HB evolutionary tracks used to determine
the position of a HB star in the H-R diagram are constructed using
the {\em modules for experiments in stellar astrophysics} code (MESA;
Paxton et al. 2011).  This stellar evolution code can pass the
helium-flash phase for low-mass stars, it is more physically complete,
and it is convenient for constructing the ZAHB models.

\begin{table}[tb]
  \small
  \centering
  \begin{minipage}[]{90mm}
    \caption[]{HB evolutionary tracks at the ZAHB.}
  \end{minipage}\\
  \begin{tabularx}{9.8cm}{XXXXXXX}
    \hline\noalign{\smallskip}
    $M/M_{\odot}$ & $M\rm_{c}/M_{\odot}$ & $M\rm_{env}/M_{\odot}$
    & log$L/L_{\odot}$ & log$T\rm_{eff}/K$ & log$R/R_{\odot}$ & log$g$  \\
    \hline\noalign{\smallskip}
    0.470    & 0.469    & 0.001  & 1.079 & 4.455 &  -0.8469 & 5.804 \\
    0.475    & 0.474    & 0.001  & 1.099 & 4.458 &  -0.8433 & 5.801 \\
    0.480    & 0.479    & 0.001  & 1.120 & 4.460 &  -0.8374 & 5.794 \\
    0.485    & 0.484    & 0.001  & 1.140 & 4.463 &  -0.8323 & 5.788 \\
    0.489    & 0.488    & 0.001  & 1.156 & 4.465 &  -0.8287 & 5.785 \\
    0.493    & 0.488    & 0.005  & 1.168 & 4.417 &  -0.7264 & 5.584 \\
    0.500    & 0.488    & 0.012  & 1.186 & 4.374 &  -0.6317 & 5.400 \\
    0.510    & 0.488    & 0.022  & 1.209 & 4.332 &  -0.5371 & 5.219 \\
    0.520    & 0.488    & 0.032  & 1.231 & 4.299 &  -0.4604 & 5.075 \\
    0.530    & 0.488    & 0.042  & 1.256 & 4.272 &  -0.3920 & 4.946 \\
    0.540    & 0.488    & 0.052  & 1.286 & 4.246 &  -0.3252 & 4.821 \\
    0.560    & 0.488    & 0.072  & 1.368 & 4.192 &  -0.1760 & 4.538 \\
    0.580    & 0.488    & 0.092  & 1.476 & 4.117 &  0.0276  & 4.146 \\
    0.600    & 0.488    & 0.112  & 1.563 & 4.026 &  0.2521  & 3.712 \\
    0.620    & 0.488    & 0.132  & 1.621 & 3.933 &  0.4679  & 3.294 \\
    0.640    & 0.488    & 0.152  & 1.660 & 3.839 &  0.6739  & 2.896 \\
    0.660    & 0.488    & 0.172  & 1.688 & 3.759 &  0.8497  & 2.558 \\
    0.680    & 0.488    & 0.192  & 1.708 & 3.737 &  0.9044  & 2.462 \\
    0.700    & 0.488    & 0.212  & 1.724 & 3.728 &  0.9305  & 2.422 \\
    0.760    & 0.488    & 0.272  & 1.758 & 3.716 &  0.9703  & 2.378 \\
    \hline\noalign{\smallskip}

  \end{tabularx}
\end{table}

As seen in Table~1, the minimum and maximum stellar
mass of the primary stars at the helium flash
is approximately 0.47$M_{\odot}$ and 0.76$M_{\odot}$,
respectively.  In other words, the stellar mass of the HB
stars produced by tidally enhanced stellar wind in our
model calculations is in the range of 0.47-0.76$M_{\odot}$;
hence, the stellar mass of the HB evolutionary
tracks constructed to determine the
positions of the primary stars on the HB in the H-R diagram
is also in the same mass range.

\begin{figure}
  \centering
  \includegraphics[width=65mm,angle=270]{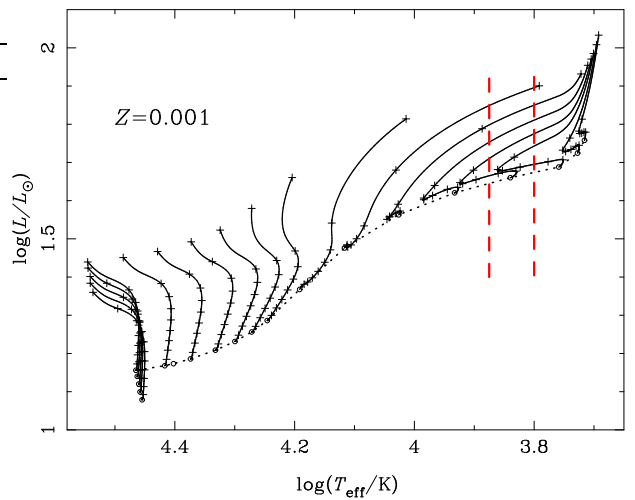}
  \begin{minipage}[]{75mm}
    \caption{Constructed HB evolutionary tracks.
     The metallicity is  $Z$=0.001.
     The HB evolution is terminated when the helium abundance drops
      below 0.001 at the star center.
      The ZAHB is shown as the dotted line at the bottom of each
       evolutionary track.
      The time interval
      between two adjacent + symbols in each track is $10^{7}$~years.
      The region between the two vertical dashed red lines denotes the
      RR Lyrae instability strip.}.
  \end{minipage}
  \label{Fig1}
\end{figure}

The information for the HB evolutionary tracks at the ZAHB
are given in Table~2.  The columns from left to  right
list the stellar mass (in units of $M_{\odot}$), the helium core
mass (in units of $M_{\odot}$), the envelope mass (in units of
$M_{\odot}$), the star's luminosity (in units of $L_{\odot}$ and
in logarithmic scale), the effective temperature, the stellar
radius (in units of $R_{\odot}$), and the logarithm of the gravity
acceleration.

The constructed HB evolutionary tracks listed in Table 2 are shown
in Fig.~1.  The helium core masses of
all tracks are about 0.488$M_{\odot}$ at the ZAHB,
except for the first four tracks in Table~2 (corresponding to
the four HB tracks that have the faintest ZAHB points in this figure).
These  four HB tracks are used to determine the HB
positions of primary stars that
undergo hot helium flash (see Section~3.1) and
have smaller helium core masses at ZAHB
than those of primary stars undergoing helium flash at the RGB tip.
The time interval between two adjacent +
symbols on each track is $10^{7}$ years.  The HB evolution is terminated
when the central helium abundance by mass fraction has dropped below
0.001.  The RR Lyrae instability strip is marked by the two vertical dashed
red lines in Fig.~1, which is used to distinguish between the blue and the red
HB stars in GCs,  and it is defined
by the vertical region: 3.8$<$log$T\rm_{eff}<$3.875 on the H-R diagram in
this paper (see Koopmann et al. 1994; Lee et al. 1990).

\subsection{Obtaining the positions of HB stars in the H-R diagram}

To obtain the
position of each primary star of our binary
samples on the HB in the H-R diagram, we
need to know $M\rm_{HF}$,  $M\rm_{c,HF}$ and the time spent on the HB.
The $M\rm_{HF}$ and  $M\rm_{c,HF}$
for each primary star
are obtained by interpolating with the results in Table 1.
The time spent on the HB is generated by a uniform
random number between 0 and the lifetime of the HB, $\tau
\rm_{HB}$.   Here,
$\tau \rm_{HB}$ is set to be the lifetime of  an HB star
with the lowest stellar mass
among the HB evolutionary tracks (i.e., 0.47$M_{\odot}$)
in Table~2, which means
that this star has the longest lifetime on the HB.  Therefore,
some of the HB stars are given a time longer than their
lifetime on the HB, and these stars are considered to have evolved
into the next evolutionary phase (e.g., AGB or WD).
This is equivalent to the scenario that RGB stars enter the HB
at a constant rate (see Lee et al. 1990).
Using $M\rm_{HF}$, $M\rm_{c,HF}$, and
the time spent in the HB phase for HB stars,
we obtain the position of each primary star on the HB
(e.g., effective temperatures and luminosities) in the
H-R diagram by interpolating among the HB evolutionary tracks shown
in Fig.1.

As seen in Table 1,  primary stars with short
orbital periods (e.g., log$P/\rm day<3.26$ in Table~1) lose
a substantial amount of envelope mass (approximately 40\%)
through the powerful tidally enhanced stellar wind.  Thus, they
leave the RGB after experiencing a huge mass loss,
and  undergo hot helium flashes at  higher temperatures (see
Section~3.1 above).  These stars have lower helium core masses
than the stars that undergo the helium flash at the tip of the
RGB (about 0.488$M_{\odot}$ in our model calculations;
see Table~1).  Therefore, in our model calculations, the core
masses of stars undergoing  hot helium flashes are in the range
between the minimum core mass required for the helium flash
(i.e., 0.4706$M_{\odot}$; see Table~1) and 0.488$M_{\odot}$.
We find, however, that the corresponding envelope masses
are nearly the same (0.001$M_{\odot}$; see Table~1).  For this
reason, the positions of the hot flash stars
on the HB in H-R diagram are obtained
by interpolating their core masses using the first five HB
evolutionary tracks given in Table~2.
We find that stars that undergo a normal  helium
flash at the RGB tip have nearly the same helium core mass of
approximately 0.488$M_{\odot}$. Therefore,   the locations of these stars
on the HB in the H-R diagram are obtained by interpolating
among the remaining HB evolutionary tracks given in
Table~2 using the envelope mass.

\begin{figure}
  \centering
  \includegraphics[width=65mm,angle=270]{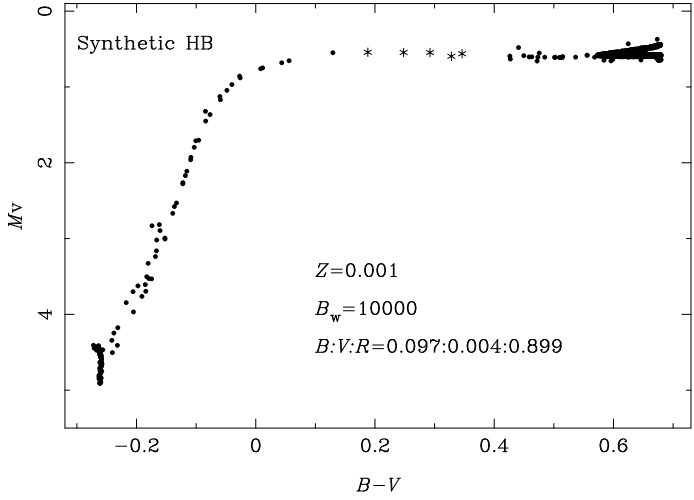}
  \begin{minipage}[]{80mm}
    \caption{Synthetic HB morphology for $Z=0.001$ and $B\rm_{w}=10^4$.
      Here, $B$, $V$, and $R$ are the numbers of HB stars located to
      the left (bluer), within, and to the right (redder) of the RR Lyrae
      instability strip.  The HB stars within the RR Lyrae
      instability strip are denoted by asterisks, whereas the other HB
      stars are denoted by dots.}. \end{minipage}
  \label{Fig2}
\end{figure}

\subsection{Synthetic HB morphology in CMD} \label{bozomath}

\begin{figure}
  \centering
  \includegraphics[width=65mm,angle=270]{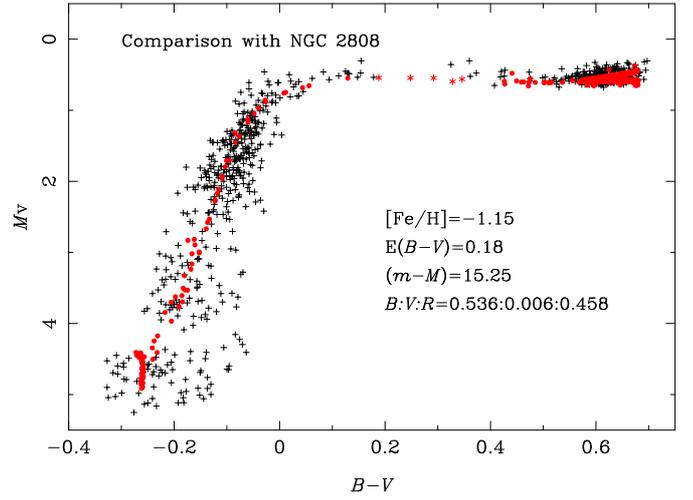}
  \begin{minipage}[]{80mm}
    \caption{Comparison of our synthetic HB with the observed
      HB morphology for globular cluster NGC 2808.  The basic
      parameters of NGC 2808 are as labeled.  The number ratio of
      stars, $B$: $V$: $R$, in different parts of the HB for this
      cluster derived from observations are also shown.  The black
      + in this figure denote the observational data of HB stars
      for this GC.  The red dots and the red asterisks denote the
      synthetic HB stars obtained from our model calculations,
      as in Fig.~2.  The photometric data for NGC 2808 are taken
      from Piotto et al. (2002).}.
  \end{minipage}
  \label{Fig3}
\end{figure}

To compare our results directly with observations,
we transformed for each primary star in our binary samples
the effective temperature and
luminosity on the HB into $B-V$ colors
and absolute magnitude, $M\rm_{v}$, respectively,
using the stellar spectra library compiled by Lejeune et al.
(1997, 1998).  Fig. 2 shows a synthetic HB in CMD under
the tidally enhanced stellar wind.
In this figure, the HB
stars located within the RR Lyrae instability strip are denoted
by asterisks, and the other HB stars are denoted by dots.
The legend in Fig.~2 shows the number ratio of stars
in different parts of the HB: $B:V:R$, where $B$, $V$,
and $R$ are the numbers of HB stars that are bluer than
(or to the left of), within, and redder than (or to the right
of) the RR Lyrae instability strip (Lee et al.
1990).

In Fig.2, our model calculations reproduce the red, blue, and extreme HB
stars. We find that about 90\% of stars shown
in Fig.~2 are red HB stars, and that about 10\% of
stars are blue HB and EHB stars.  There are only a few stars
located within the RR Lyrae instability strip.
The results shown in the figure demonstrate that we
are capable of reproducing the basic HB morphology by
considering tidally enhanced stellar winds only, and without
making {\em any} assumption of mass-loss dispersion
on the RGB.  This result is consistent with our expectation that
different initial orbital periods of the binaries yield different
mass losses on the RGB, and hence different positions on the
HB.

\subsection{Model comparison with observations for GC
  NGC 2808} \label{bozomath}

NGC 2808 is a typical GC for which the 2P problem exists, which
is why it has been studied extensively by various researchers.
The cluster has an intermediate metallicity of about $Z=0.0014$
(or [Fe/H]= -1.15; Harris, 1996), and its age is in the range of
10.4-12.9~Gyr (Gratton et al. 2010).  The cluster, however,
exhibits a bimodal HB morphology (i.e., red, blue, and extreme
HB are well populated, with a few stars in the RR Lyrae instability strip;
Bedin et al. 2000; Piotto et al. 2002).  Over the years, several
2P candidates have been proposed for this cluster (e.g., age,
helium enhancement, etc. D'Antona \& Caloi, 2004, 2008).
None of them alone, however, can successfully
explain the HB morphology of NGC 2808.

Since the metallcity and the age used in our model calculations are
very similar to those for NGC 2808, we can directly compare
between our synthetic HB  and the observed HB
in this cluster.  The results of this comparison are shown in Fig.~3.
The photometric data for NGC 2808 adopted in our study were
obtained by Piotto et al. (2002) with the HST/WFPC2 camera
in the $F439W$ and $F555W$ bands.  The distance module
for the cluster is 15.25 and the reddening correction is 0.18
(Bedin et al. 2000; D'Antona \& Caloi, 2004).  In the
observational data of Piotto et al. (2002), about 46\% of the HB
stars in the cluster are red HB stars, and the other 54\% of the HB
stars are located in the region that is bluer than the RR Lyrae instability
strip.

In Fig.3, we combine our synthetic HB with the observed HB
morphology for globular cluster NGC 2808.  The black +
symbols correspond to the observed HB stars for the cluster.
The red dots and the red asterisks denote the synthetic HB
stars, as in Fig.~2.  Fig.~3 demonstrates that by considering
{\em only} tidally enhanced stellar winds as the 2P, we can
successfully reproduce the basic HB morphology of NGC 2808
(without making any assumption for the mass-loss dispersion
on the RGB at all).  Red, blue, and extreme HB are present both
in the synthetic HB and in the observed one for NGC 2808 \footnote
{One can see in Fig.~3 that the blue HB stars present an enormous
spread of magnitude at a given color (e.g., the scatter in $M\rm_{V}$
appears to be about 3 magnitudes) in this cluster.
This is due to not only the decreasing
sensitivity of B-V colors to temperatures but also
to the increasing bolometric correction
for hotter stars, i.e., the maximum of stellar flux
is radiated at ever shorter wavelengths for
increasing temperatures, making stars fainter at V
(Moehler 2010).}.
Furthermore, we successfully reproduce the gap between the
red and the blue HB stars for the cluster.

We find, however, that the number ratio of stars for different
parts of the HB, $B:V:R$, is very different for the
synthetic HB and the real HB for NGC 2808.  It turns out that
only about 10\% of the HB stars are located in the blue and
the extreme HB  (as opposed to 54\% in the real data), and
about 90\% of the HB stars are red HB stars in our synthetic
HB results (versus 46\% in the real data).  This means that
the HB stars in the region bluer than the RR Lyare instability
strip are underrepresented in our model calculations.
Dynamical interactions could partly alleviate this problem.
Globular clusters provide a good environment
for dynamical interactions due to their high density.
These interactions may enhance the
mass loss of the HB progenitors, e.g., by close encounters of red giant
stars with main-sequence stars or compact objects (Moehler,
2010). Some observations seem to support this.
Fusi pecci et al. (1993) claimed that more concentrated or more dense
clusters have a bluer and longer HB morphology.
Buonanno et al. (1997) demonstrated that the higher
the central density of the cluster, the higher the relative
number of stars that populate the most bluest region of the HB.
However, the exact dynamical mechanism for the enhancement of mass loss
remains unclear.   Furthermore, we do not know
the real initial orbital-period distribution for binary
systems in GCs at present, which may also impact our model
calculations results. Investigating the impact of these effects,
however, was beyond the scope of the present work, and it
will be the subject of a follow-up study.

In Fig.~3, one can see that our synthetic EHB stars cannot reach
the faintest HB stars observed in NGC 2808.  The latter are
blue-hook stars, as discussed in Section 3.1, which correspond
to a higher effective temperature than the canonical EHB stars.
Brown et al. (2001) have suggested that these stars undergo a
late hot helium flash while descending along one of the WD
cooling curves.  The helium and carbon abundance in the
envelope of these stars are enhanced by the helium flash mixing,
which is not considered in our model calculations (see discussion
in Section 3.1).  That is the main reason why our synthetic EHB
stars cannot reach the faintest HB stars observed in NGC 2808.
Our model results, however, demonstrate that tidally enhanced
stellar winds during binary evolution can explain the anomalous
mass loss while on the RGB, which is required to explain the
existence of blue-hook stars in GCs.

\section{Discussion}

We have proposed that tidally enhanced stellar wind
in binary evolution may influence the
HB morphology in GCs, and we showed that we can produce
the red, blue, and extreme HB stars without
any additional assumption of mass-loss dispersion on
the RGB. This is the largest advantage that distinguish our models
from other 2P candidates that were based on
the single-star evolution.
Although recent observations revealed that the fraction of binary systems in
GCs (Sollima, et al. 2007; Dalessandro, et al. 2011) is low compared
to the population of isolated stars,   this does not mean
that binaries are not important for the evolution of GCs.  On the
contrary, it is well known, for example, that dynamical interactions
in GCs can destroy binary systems; i.e., soft binaries can be easily
disrupted by any strong encounters.  The binary evolution
can also lead to their disruption (e.g., supernova explosions,
star mergers, etc.).  Furthermore, Ivanova et al. (2005) have suggested
that, to explain the currently observed low fraction of
binary systems in cluster cores, the initial ratio of stars in binaries
to all stars in GCs should be very high (even close to 100\%).  This
means that the population of binary systems is much higher when
a GC is formed, and that binaries may play an important role in
the evolution of GCs.

To investigate the effects of the tidal enhancement parameter,
$B\rm_{w}$, on the HB morphology, we reduced this parameter 20 fold,
and the corresponding results for the synthetic HB CMD are shown in
Fig.~4. All other model parameters remain the same as in
Fig.~2.  By comparing Fig.~2 with Fig.~4, one can see that although
$B\rm_{w}$ is 20 times different between the two cases, the corresponding HB
morphology is nearly identical for the two cases: the red, blue, and
extreme HB stars are clearly reproduced in both cases, and the number
ratio of stars at different parts of the HB is very similar.  As in
Fig.~2, in Fig.~4 we find that about 91\% of the HB stars are red
HB stars, and that 8\% of the HB stars are blue HB and EHB stars.
Furthermore, in both cases there are only a few HB stars located
within the RR Lyrae instability strip.  This comparison demonstrates
that the HB morphology is {\em not} very sensitive to the tidal
enhancement parameter, $B\rm_{w}$.

The mass ratio of primary-to-secondary in this
work is 1.6, but we also adopted several different mass ratios
of primary-to-secondary (namely, 1.2, 1.6, and 2.5) in our
model calculations, but we found that this mass
ratio has little influence on our final results.

The metallicity in our model calculations is $Z$=0.001, which is
a typical value for GCs.  We also performed calculations for metallicity value 20
times higher (i.e., $Z$=0.02) to investigate
its impact on the results.  We found that even for this high value of the
metallicity, the blue and extreme HB stars can still be produced.  That
is because the initial orbital periods of some binary systems are short,
and so the primary stars in these binaries can lose much of their
envelope mass due to tidally enhanced stellar winds.  Thus, our
model can be applied to explain the formation of extended HBs,
which are found in some metal-rich GCs, such as NGC 6388 and NGC 6441
(Rich et al. 1997).

\section{Conclusions}

\begin{figure}
  \centering
  \includegraphics[width=65mm,angle=270]{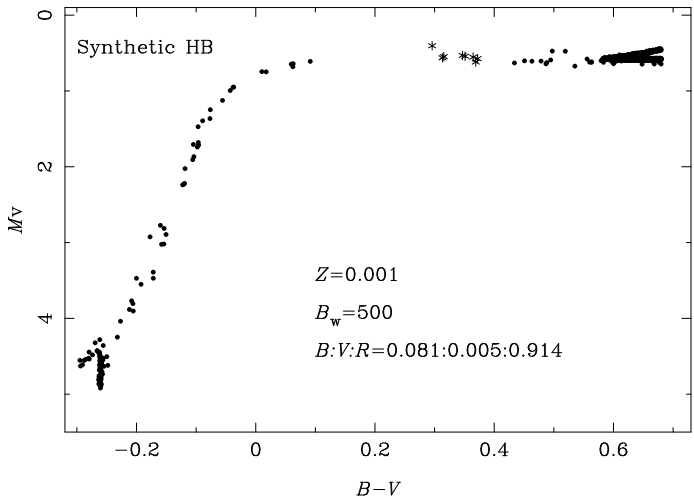}
  \begin{minipage}[]{80mm}
    \caption{The same as that shown in Fig.~2, but for  $B\rm_{w}$=500. }.
  \end{minipage}
  \label{Fig4}
\end{figure}

The purpose of the present study was to investigate the effects of
tidally enhanced stellar winds during the binary evolution
on the HB morphology of GCs.  We  found
that, with the tidally enhanced stellar wind,
red, blue, and extreme HB stars are
all produced in our synthetic HB,
and the HB morphology is {\em not} very sensitive to the tidal
enhancement parameter, $B\rm_{w}$.
Furthermore, we compared the synthetic HB with the observed one in
the GC NGC 2808 .  We  found that the basic
HB morphology of NGC 2808 can be
reproduced well in our models,
which use only the tidally enhanced stellar winds
as the 2P, and without making any assumption on the
mass-loss dispersion on the RGB.  Moreover, we
successfully reproduced the gap between the red and the blue
HB stars for this cluster. However,  the
number ratio of stars for different parts of the HB, namely
$B:V:R$, is very different between our synthetic HB and the observed
HB in NGC 2808: only about 10\% of the HB stars are
located in the blue and the extreme HB  (as opposed to
54\% in the real data), and about 90\% of the HB stars are
red HB stars in our synthetic HB (versus 46\% in the
real data).  This implies that fewer HB stars in the region bluer
than the RR Lyrae instability strip are produced in our
model calculations compared to the real one in NGC 2808.

\begin{acknowledgements} It is a pleasure to thank the referee,
Professor Peter Eggleton for
his valuable suggestions and comments, which improved the paper
greatly. The authors would like to thank  Ilia Roussev
for his constructive comments made towards improving the quality
of this manuscript. This work is supported by the National
Natural Science Foundation of China (Grant No. 11033008, 10973036, 11173055 and
11273053) and the Chinese Academy of Sciences (Grant No.
KJCX2-YW-T24).

\end{acknowledgements}

\clearpage

\end{document}